\documentclass[conference]{IEEEtran}
\IEEEoverridecommandlockouts
\usepackage{cite}
\usepackage{amsmath,amssymb,amsfonts}
\usepackage{algorithmic}
\usepackage{graphicx}
\graphicspath{ {./images/} }
\usepackage{textcomp}
\usepackage{xcolor}
\def\BibTeX{{\rm B\kern-.05em{\sc i\kern-.025em b}\kern-.08em
    T\kern-.1667em\lower.7ex\hbox{E}\kern-.125emX}}
\begin{document}

\title{Literature Review on Image Compression, Tracking, Adaptive Training and 3D Data Transmission\\
}

\author{\IEEEauthorblockN{Rajat Bothra Jain}
\IEEEauthorblockA{\textit{Department of Computing Science}\\
\textit{University of Alberta}\\
Edmonton, Canada\\
bothraja@ualberta.ca} 
\and
\IEEEauthorblockN{Sravanti Chinta}
\IEEEauthorblockA{\textit{Department of Computing Science} \\
\textit{University of Alberta}\\
Edmonton, Canada\\
sravanti@ualberta.ca}
}

\maketitle

\begin{abstract}
The literature review presented below on Image Compression, Transmission of 3D data over wireless networks and tracking of objects is the in depth study of Research Papers done in Multimedia lab. Most of the papers presented in this literature review have tackled the problems present in the conventional system and offered an optimal and practical solution.
\end{abstract}

\section{Gaussian and Laplacian of Gaussian weighting functions for robust feature based tracking} 

\subsection{Summary}
\begin{figure}
    \centering
    \includegraphics[width=0.5\textwidth]{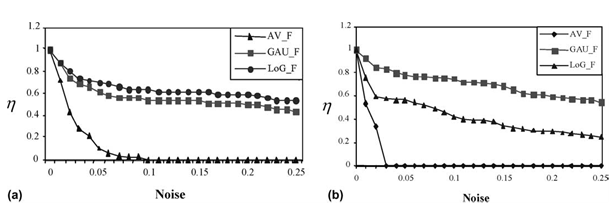}
    \caption{ (a) Performance of weighting functions with synthetic car sequence, 50 features have been initialized and are being tracked. (b) Performance of weighting functions with real image car sequence, 50 features have been initialized and are being tracked.}
    \label{fig:noise}
\end{figure}

In this paper \cite{singh2005gaussian} the author  addresses the effects of noise on the performance of the standard KLT algorithm and proposes two solutions - Laplacian of Gaussian (LoG) and Gaussian weighting function to increase immunity of the tracking algorithm to effects such as poor CCD exposure and compression noise. The KLT algorithm is limited by its use of averaging as a weighting function over the feature window. KLT algorithm fails to track the object even at low levels of noise, while the weighted tracking algorithm continues to track until 25\% noise. Fig. \ref{fig:noise} shows the affect of noise on performance tracking algorithm.

\subsection{Analysis}
Irrespective of the input test sequences, both the proposed weighting functions, Gaussian weighting function and LoG weighting function, consistently perform better tracking in a random noise environment than the KLT average weighting function. Real world tracking sequences are expected to be noisy, and therefore are ideal for application of these weighting functions.The LoG  is suitable for images with well-defined and sharp image corners like the synthetic test images. Real images where the edges may not be as sharply defined, the Gaussian weighting function performs better.

\section{Integrating active face tracking with model based coding} 

\subsection{Summary}
In this paper \cite{yin1999integrating} the author  discusses the tracking of a talking face with an active camera whereas most previous work done focuses on the detection on a still camera. The detection and tracking of the active talking face is a fundamental step towards realizing an application of MPEG4 in real situations. First, the background compensation in successive frames is done, and the motion-energy tracking approach is used coupled with a morphological filter to reduce the noise. Third, the facial features are detected using Hough Transform and deformable template coupled with color information. Finally, a wireframe model is adapted to the extracted face using a coarse-to-fine adaptation algorithm.

\subsection{Analysis}
Proposed a system for tracking a face observed with an active camera for an arbitrary background. Camera mounted on an active platform (pan/tilt) is used to take an active video sequence, which shows a talking person with an unconstrained background Initial results show that this approach is feasible in practical applications.

\begin{figure}
    \centering
    \includegraphics[width=0.5\textwidth]{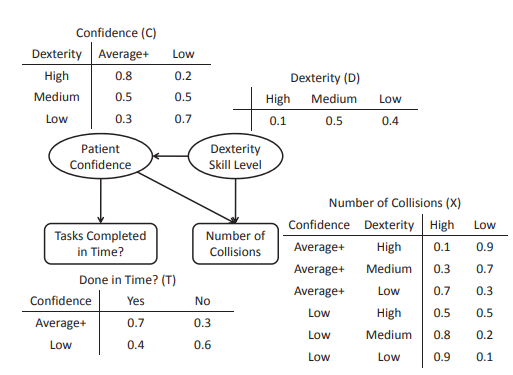}
    \caption{A sample Bayesian network topology}
    \label{fig:bayesian}
\end{figure}

\section{A Framework for Adaptive Training and Games in Virtual Reality Rehabilitation Environments} 

\subsection{Summary}
In this paper \cite{rossol2011framework}, the author addresses about Bayesian networks as a way to measure student performance in online web-based multimedia educational games but which have not been used in Virtual Reality Rehabilitation. Fig. \ref{fig:bayesian} shows the Bayesian network topology. This paper proposed a novel framework based on Bayesian networks for self-adjusting adaptive training in virtual rehabilitation environments. The system offers standard training modes and interactive virtual reality games for increased patient engagement. In addition, the system provides a framework to allow clinicians to design, build, and customize interactive indoor rehabilitation training environments that can dynamically respond to user actions in a meaningful way.

\subsection{Analysis}
Participants who used the proposed virtual reality wheelchair training system completed the real world obstacle course on average faster, with a mean time of 81.5 seconds for the experimental group compared to 104.5 seconds for the control group .

\section{Interactive Multimedia for Adaptive Online Education}

\subsection{Summary}
Through computer reinforcement, the author \cite{cheng2009interactive} addresses a vision of providing publicly accessible education anywhere, at any time, and to anyone. Online Multimedia Education (CROME) combines the primary components of education, such as learning, teaching, and testing, as well as adaptive testing and student modelling. Online tutoring addresses issues such as time difference, curriculum diversity, and potential language and cultural barriers. The use of digital multimedia and e learning is a more widely adopted and effective approach to promoting global education. We interpret a molecular structure as a graph, where nodes are atoms and edges are bonds, to evaluate the correctness of an answer and award partial marks.The traditional use of multimedia content in education emphasises subject knowledge but not cognitive skills. Testing students' broader range of skills is now possible thanks to the use of multimedia items.The traditional use of multimedia content in education emphasises subject knowledge but not cognitive skills.

\subsection{Analysis}
The CROME aims for portability, resilience, scalability, and interoperability. The current implementation lays the groundwork for the development of a more robust, effective, and far-reaching multimedia architecture. Using the framework as a starting point, this paper aims to raise awareness among the multimedia research community and to encourage participation in topics that have received insufficient attention.

\section{Nose Shape Estimation and tracking for model based coding} 

\subsection{Summary}
 This paper \cite{yin2001nose} addresses the importance of nose shape, which is an important par of facial expression recognition and generation.Face feature extraction is critical in model-based coding and human face recognition applications. Detecting and tracking the nose shape is not easy, and it is just as important as the eyes and mouth for model-based coding, particularly for the analysis and synthesis of realistic facial expressions. Experiments on real video sequences for low bit rate video coding show the benefit of the proposed scheme. Individual nostril and node templates have been created. To reduce the sensitivity of facial feature detection to noise, a two-stage region growing algorithm - global region growing and local region growing - is developed, with a large growing threshold chosen to explore more skin area.The feature regions are extracted by segmenting the image using region growing on the colour components, namely luminance (Y) and chrominance (C) (U,V). Following the collection of feature blobs, they are classified using the k means classifier. To correctly detect the nostril shape, a geometric template, which is a twisted pair curve with a leaf-like shape, is applied to the nostril region.

\subsection{Analysis}
The new method restricts the search region and employs the shape adaptive model to compensate for the shrinking effect of deformable templates. The deformable template matching algorithm fits the template to the facial features in a few single steps, avoiding the computational complexity of a large number of parameter searches and the template shrinking problem. The results demonstrate that the approach addressed in this paper is applicable in real-world applications.
\section{Visual gesture recognition for ground air traffic control using the Radon transform}

\subsection{Summary}
\begin{figure}
    \centering
    \includegraphics[width=0.4\textwidth]{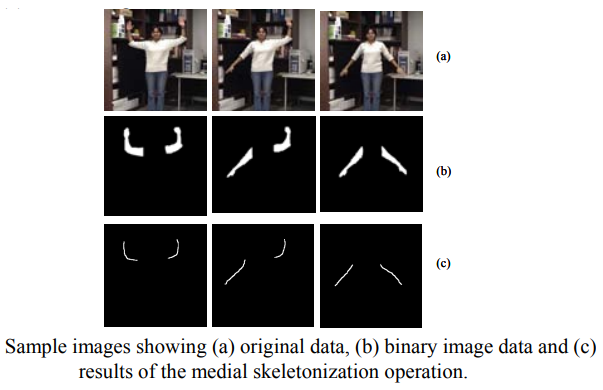}
    \caption{}
    \label{fig:gesture2}
\end{figure}

In this paper\cite{singh2005visual} the author has discussed a novel method for the recognition of hand gestures, used by air marshals for steering aircraft on the runway, using the Radon transform. Foreground is extracted from background by using an adaptive threshold value. Thinning is applied to images to obtain the medial skeleton. Fig. \ref{fig:gesture2} shows the medial skeleton of each segment. Radon Transform is used to detect orientation of skeletal lines. Once the Radon transform has been computed, we threshold the Radon transform coefficients to extract the most significant local regions. It is then normalized between 0-1. Finally, a k means clustering with k=8 is applied for classification.

\subsection{Analysis}
The proposed approach discussed by the author in this paper \cite{singh2005visual} does not require any normalization of the image itself and promising recognition rates achieved with the parameterized Radon transform. However, it is observed that due to poor segmentation and thinning the medial skeleton is no longer a unitary segment but has loop like regions. These regions act as noise sources and cause a larger than usual Radon transformation. Thus when the Radon transform is thresholded these noise sources are not eliminated and get incorporated into the cumulative Radon transform and subsequently the normalized cumRT. These, noisy feature vectors sometimes lead to misclassification.

\section{Variable Resolution Teleconferencing} 

\subsection{Summary}
The author in this paper \cite{basu1993variable} describes an image compression scheme based on Variable Resolution (VR) sensing. Compressed methods can be broadly classified as lossless or lossy, depending on how the method affects the data being compressed. The compression of motion pictures can be used in teleconferencing(video phones), electronic storage of movies , or even television transmission.The concept of variable resolution (VR) has been used in a variety of fields, but its application to image compression is novel. Images compressed with the VR transform can be compressed further using other methods. A lookup table can be used to perform the VR transform. Any entropy technique's ability to compress data is dependent on the presence of patterns in the input data stream. Any entropy encoding technique's ability to compress data is dependent on the presence of patterns like repeating bytes or skewed frequency distributions, which digitized computer images lack.The VR transform has two parameters that influence the final image: the expected savings (compression) and alpha, which controls the distortion at the image's edges in relation to the fovea. VR cannot be used to compress all data because the distortions caused in many cases (such as commercial television broadcasts) would be unacceptable.

\subsection{Analysis}
As an image compression technique , variable resolution has undeniable advantages. The rate at which it can compress images, particularly on slower machines, and the high quality present in the foveal region make it ideal for the teleconferencing market.

\section{Stereo Matching Using Random Walks} 

\subsection{Summary}
This paper \cite{shen2008stereo} introduces a new two phase stereo matching algorithm based on the random walks framework, with promising results when tested on the Middle bury test bed. Stereo matching is a critical and traditional task in stereo vision, with the goal of determining the disparities of corresponding pixels in a pair of stereo images. A recent application of random walks in multi label image segmentation demonstrates great promise for its use in stereo matching. Prior matrices defined on the penalties of different disparity configurations and Laplacian matrices defined on pixel neighbourhood information are used to extract a set of reliable matching pixels. Following that, using the reliable set as seeds, the disparities of unreliable regions are calculated by solving a Dirichlet Problem. The algorithm operates in the (add that) space, and the parameters are automatically adjusted based on the lighting conditions of the input image pair. The main disadvantage highlighted in this paper is that only an exact and unique minimum solution of an energy function of the above form can be produced, whereas other approaches, such as Graph Cuts(GC), can only produce an approximation.
\subsection{Analysis}
The algorithm was implemented in Mat Lab, and at least a tenfold speedup is expected once the author's \cite{shen2008stereo}  algorithm is implemented in GPU, which will be significantly faster than many other global methods using a different optimization framework. The main advantage of random walks is that they can produce an exact and unique minimum solution of an energy function of the above, whereas other approaches, such as GC, can only produce an approximation. This paper not only proposed a new algorithm, but also demonstrated the feasibility of a new framework for solving stereo matching, random walks . The algorithm, which operates in the (add that formula) space, is limited to colour images.

\section{Eye Tracking and Animation for MPEG-4 Coding} 

\subsection{Summary}
The author \cite{bernogger1998eye} of this research describes a method for synthesising eye movement by computing the deformation of the 3D model's eyes using the retrieved eye attributes. Create a 3D face model that is unique and map it to the first frame of the face sequence. The deformation parameters were used to the 3D model based on the retrieved eye attributes to synthesise the eye movement in successive frames. Determine two coarse zones of interest for the eyes in the eye tracking and detection algorithm. Using a gradient-based Hough transform, search the iris of the eyes. Select a small area of interest for obtaining the eye borders. Getting an initial approximation for the eye lids using colour information. Localize the eye lids using deformable templates.The eye detection and tracking algorithms extract the contours of the iris and eye lids in each frame of the image sequence. These eye features are used to synthesise the real motions of the eye on a 3D facial model.

\subsection{Analysis}
The author in this paper\cite{bernogger1998eye} found from experiments that the Hough transform, which is the first and therefore one of the most important steps, is very robust against noise as well as edges which are not produced by the contour of the iris. A high resolution facial model with 3118 patches is used to evaluate the algorithm. The proposed algorithms behaves well for synthesizing eye movements. 

\section{Perceptually Optimized 3D Transmission Over Wireless Networks} 

\subsection{Summary}
The essential factors determining the display quality of three-dimensional objects (3D) are reviewed in this paper \cite{cheng2007perceptually}. The ability to adaptively adjust the model representation while maintaining satisfactory quality as perceived by a viewer is an important consideration in designing effective interactive online 3D systems. An overview of the research on developing a 3D perceptual quality metric that integrates two important ones, texture resolution and mesh resolution, which control the transmission bandwidth requirements. Simplification algorithms attempt to control mesh complexity by developing various strategies for simplifying the Level of Detail(LOD) in various parts of a 3-D object. One of the major drawbacks of most 3D transmission algorithms is that they do not account for data loss. Despite the fact that issues of multimedia transmission over wireless networks have received attention, little work has been done on wireless 3D transmission. Some preliminary implementations for deploying 3POPIT(3D Perceptually Optimized Partial Information Transmission) over a lossy wireless network are highlighted in the paper. The two programs shown are combining and interpolating based on different texture and mesh sub samples, as well as a comparison of perpetually optimised versus non optimised transmission.
\subsection{Analysis}
In this paper\cite{cheng2007perceptually}, the author  has discussed the factors that influence 3D image degradation and proposed a method for estimating perceptual quality while taking mesh and texture resolution variations into account. Following that, a strategy for optimising perceptual quality in the face of packet loss was presented. The approach is validated by the experimental results. Given partial packet transmission, the most computationally efficient approach to interpolation would be to predetermine neighbours and coefficients for interpolation and store various look up tables. Packet size and header length are not considered in the packet loss model, which is a simple model intended to demonstrate the feasibility of the method depicted.

\bibliographystyle{IEEEtran}
\bibliography{bibliography.bib}
\end{document}